# Reducing Total Power Consumption Method in Cloud Computing Environments

Shin-ichi Kuribayashi[1]

[1]Department of Computer and Information Science, Seikei University, Japan

E-mail: kuribayashi@.st.seikei.ac.jp

*Abstract*

*The widespread use of cloud computing services is expected to increase the power consumed by ICT equipment in cloud computing environments rapidly.*

*This paper first identifies the need of the collaboration among servers, the communication network and the power network, in order to reduce the total power consumption by the entire ICT equipment in cloud computing environments. Fivefundamental policies for the collaboration are proposedand the algorithm to realize each collaboration policy is outlined.Next, this paper proposes possible signaling sequences to exchangeinformation on power consumption between network and servers, in order to realize the proposed collaboration policy.  Then, in order to reduce the power consumption by the network, this paper proposes amethod of estimating the volume of power consumption by all network devices simply and assigning it to an individual user.*

## 1. Introduction

Global warming is a top-priority issue requiring an urgent, global-scale response. There is a high hope that the use of ICT will dramatically reduce power consumption, for example, by optimizing the collection and delivery of goods and reducing the amount of travel[1],[2]. However, the widespread use of ICT equipment is expected to increase the power consumed by ICT equipment rapidly and it is recognized that the power consumption of ICT equipment themselves should be one of key issues.

Cloud computing services are rapidly gaining in popularity[3]-[7].They allow the user to rent, only at the time when needed,only a desired amount of computing resources (processingability and storage capacity) out of a huge mass of distributedcomputing resources without worrying about the locations orinternal structures of these resources. It isanticipated that enterprises will accelerate their migrationfrom building and owning their own systems to renting cloudcomputing services because cloud computing services are easyto use, and can reduce both business costs and environmentalloads.  The cloud computing environments require a huge amount of ICT equipment such as servers, storage devices, communication network devices and client terminals. Therefore, it is clear that the widespread use of cloud computing services will greatly contribute to a rapid increase on ICT power consumption.

Most conventional measures to save the power consumed by servers and the communication network have been discussed and implemented independently [8]-[19]. For example, slowing the processing of a server to reduce its power consumption can prolong not only its processing time but also the bandwidth holding time in the network, which in turn increases the power consumed by the network. Conversely, raising the processing speed of a server increases its power consumption but reduces the processing time, and consequently reduces the power consumed by the network. This may reduce the total power consumption. Therefore, it is important to take an integrated approach to saving the power consumed by both servers and the network.

To reduce the power consumed by the network, it is necessary to measure the power consumed by all network devices accurately and simply. One method proposed to estimate the total network power consumption from the volume of shipment of network devices. It is necessary to estimate power consumption more accurately if the emission of carbon dioxide should be calculated from power consumption of each network user. However, it is not realistic to seek to find carbon footprint for each packet in the same way that the transportation system finds a carbon footprint for each package. It is necessary to consider a simpler way of estimating the power consumed by an individual network.

Moreover, it is expected that the percentage of power generated by renewable power, such as photovoltaic power generation and wind power will continue to rise. Since the power generated by these natural means varies over time considerably, the electric power storage (high-capacity batteries) needs to be introduced to stabilize power supply [19],[20]. This implies that it is necessary in the future to consider cases where ICT equipment operates under the condition that the total power supply available is restricted.Therefore, it is required to take the restriction of the available power supply into consideration.

The rest of this paper is organized as follows. Section 2 explains the system model for cloud computing environments and identifies fundamental policies that should be adopted for the three parties - servers, the communication network, and the power network - to collaborate with each other to reduce the total power consumed by the entire ICT equipment in cloud computing environments.The algorithm to realize each policy is also outlined.Section 4proposes possible signaling sequences to exchange information on power consumption between network and servers, in order to realize the proposed policy in Section 3. Section 5 proposes a method of estimating the volume of power consumption by all network devices simply and assigning it to an individual user, in order to reduce the power consumption by the network.Finally, Section 6 presents the conclusions. This paper is an extension of the study inReferences [22]and [23].

## 2. Fundamental policies for reducing total power consumptionin cloud computing environments

### 2.1 Proposed system model for cloud computing environments

Figure 1 illustrates the system model for cloud computing environments, in which there are multiple areas at different locations and each area has a resource set of servers (processing ability), bandwidth which connects the selected server to the client and electric power storage which provide the power to ICT equipment in its area. Electric power storagewith limited electric capacity is treated as the power network. Here, the power restriction to ICT equipment, which could happen in the future, is taken into consideration.When a request occurs, one area is selected from among k areas, and required amounts of processing ability,bandwidth and power capacity are simultaneously allocated in the selected area.

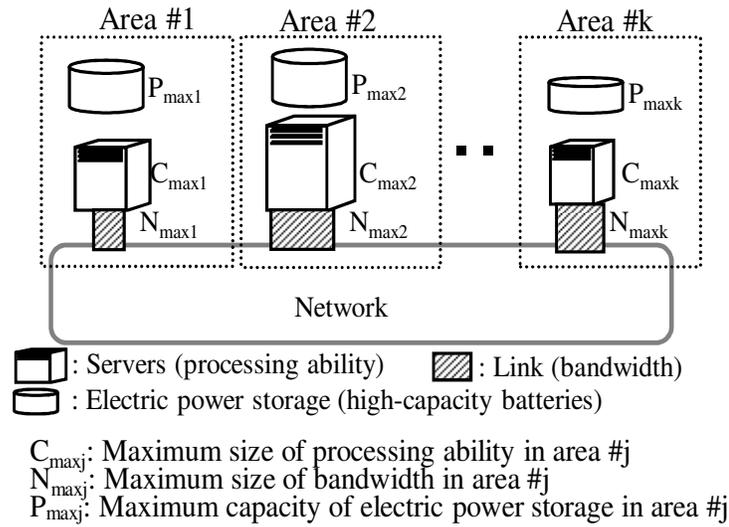

Figure 1. System model for cloud computing environments

## 2.2 Fundamental policies for reducing total power consumption

### 2.2.1 Need of the collaboration among servers, the communication network and the power network

Case 1 in Figure 2 shows an example in which the line speed is decreased to reduce the power consumed by the network. While the power consumed by the network is reduced, the lower line speed increases the time it takes for servers to transfer data, making it likely that thepower consumed by servers will increase. This may increase the total power consumption. Conversely, case 2 in Figure 2 shows an example of reducing the power consumption of servers by slowing its processing clock. While the power consumed by servers decreases, the processing time increases, which can in turn increase the time during which a bandwidth in the network is retained, and consequently the power consumed by the network increases. Therefore, it is necessary to consider integrated measures involving servers and multiple networks(**Policy** I). For simplicity, this paper assumes asingle network.

The conditions for measures to reduce power consumption by the server or the network to be effective areas below. Let $X_1$ be the total power saved by a power consumption reduction measure taken for the network, and $Y_1$ the total additional power consumed as a result by servers and client terminals affected by this measure.Any measure to reduce the power consumed by the networkshould be taken only when$X_1>Y_1$ .Similarly, let $X_2$ be the total power saved by a power consumption reduction measure taken for servers, and $Y_2$ the total additional power consumed by the network as a result. Any measure to reduce the power consumedby servers should be taken only when$X_2> Y_2$.

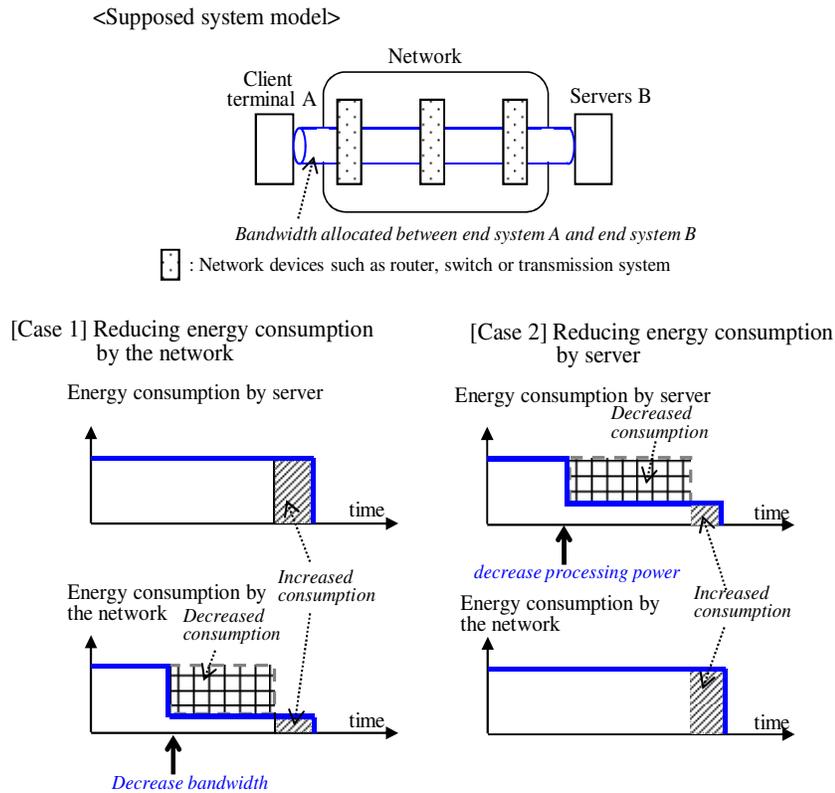

Figure 2. Examples of issues when the collaborative reduction of energy consumption is not implemented

The other possible measure is to increase the power consumption on the one side but may save more power on the other side. For example, raising the processor clock frequency of the server, in turn, reduces the bandwidth holding time and consequently reduces the power consumed by the network (Figure 3). This could decrease the total power consumption by both servers and the network.

### 2.2.2 Collaboration between servers and communication network

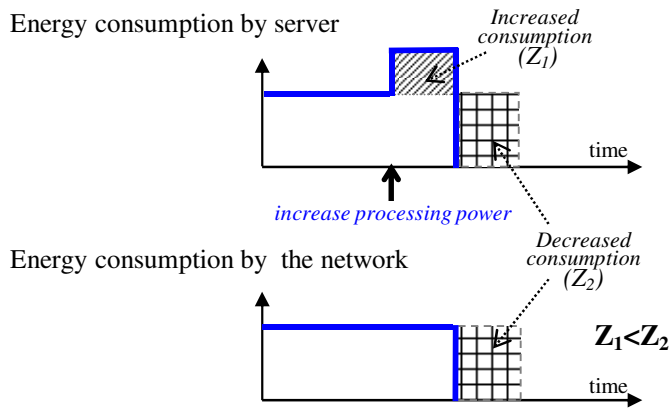

Figure 3. Example of collaborative reduction of total energy consumption

International Journal of Computer Networks & Communications (IJCNC) Vol.4, No.2, March 2012

[1] A measure that has already been proposed to reduce power consumption includes assembling as many servers and network devices as possible in some areas and keeping many of them in the sleep state for a long time. However, if devices are aggregated independently, mismatches between servers and the communication network can occur, as shown in Figure 4-(1).  As a result, the number of servers or network devices in the sleep state is reduced, or it becomes impossible to provide services in the worst case. To avoid such a situation, it is necessary to aggregate servers and network devices simultaneously as much as possible (**Policy II**), as shown in Figure 4-(2). This increases the number of servers and network devices that can be put in the sleep state, resulting in a drastic reduction in the total power consumption. The possible aggregation algorithm according to Policy II is outlinedas follows:

   <Step 1> First, we focus on processing ability.

  1) Area Xc which used the largest amount of processing ability and area Yc which used the least amount of processing ability are selected from among $m_1$ areas.  $m_1$ is the number of areas which are not in sleep mode. The algorithm will be stopped if both Xc and Yc could not be selected.

  2) If all requests currently processed in area Yc could be migrated to area Xc (both processing ability and bandwidth should be migrated), area Yc will get into sleep mode. Then go back to 1). Otherwise, requests will not be migrated and stop the algorithm.

   <Step 2> Then, we focus on bandwidth, assuming that the aggregation in step 1 is not executed.

  1) Area Xb which used the largest amount of bandwidth and area Yb which used the least amount of bandwidth are selected from among $m_1$ areas. The algorithm will be stopped if both Xb and Yb could not be selected.

  2) If all requests currently processed in area Yb could be migrated to area Xb (both processing ability and bandwidth should be migrated), all requests currently processed in area Yb are migrated to area Xb and area Yb gets into sleep mode. Otherwise, requests will not be migrated and stop the algorithm.

   <Step 3> If the final number of sleep areas in step 1 is equal or more than that in step 2, the aggregation executed in step1 is adopted. Otherwise, the aggregation executed in step 2 is adopted.

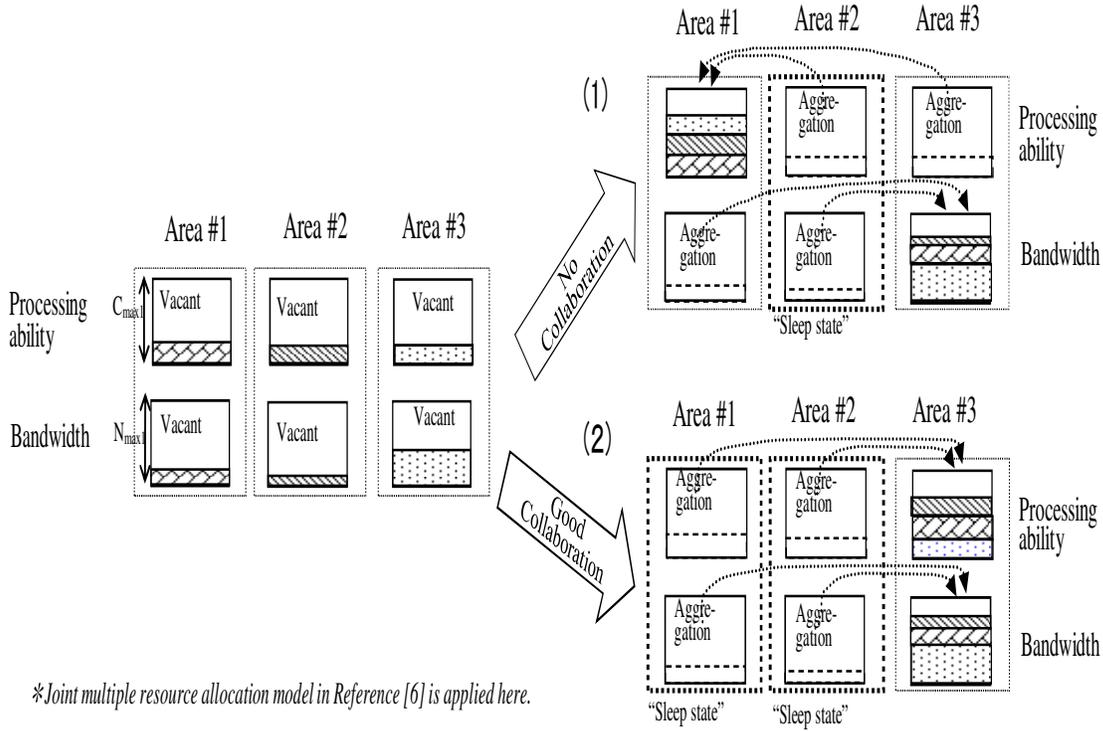

Figure 4. Collaboration image between servers and communication network

Figure 5 illustrates an example of request migration according to the algorithm explained above.There are three areas in the figure. For simplicity, the figure shows only processing ability. First, Area 1 becomes area Xc, and area 3 becomes area Yc. Next, area 2 becomes area Xc, and area 3 becomes area Yc.Finally, all requests is aggregated in area 3, and hence devices in areas 1 and 2 go to sleep mode.

[2] Preventing a part of the servers or network devices in the sleep mode from going back to the normal mode can be equivalent to saving the power that might be consumed by these devices if they were actually put in the normal mode. Figure 6 illustrates an example. It is assumed that server 1-1 (sleep mode) at area #1 needs to be put in the normal mode when a new request is generated. If it is put in the normal mode, power consumption may rise significantly. In this case, it is better that a new request be allocated to area #2 which still has a spare processing ability. This prevents the significant rise in power consumption. If area #2 does not have enough processing ability to handle a new request, server 1-1 at area#1 will be put in the normal mode.Therefore, it is necessary to take any action to keep or put servers or network devices in the sleep mode as much as possible (**PolicyIII**).

### 2.2.3 Collaboration among servers& communication network and power network

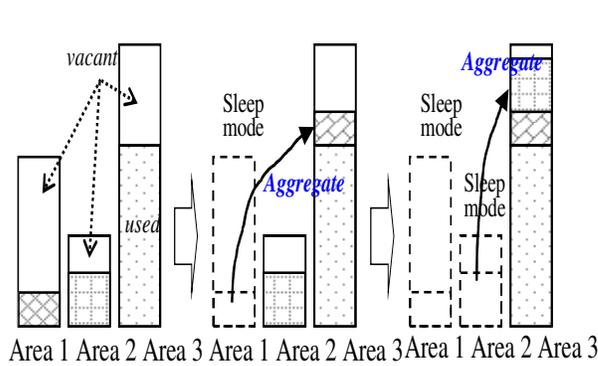
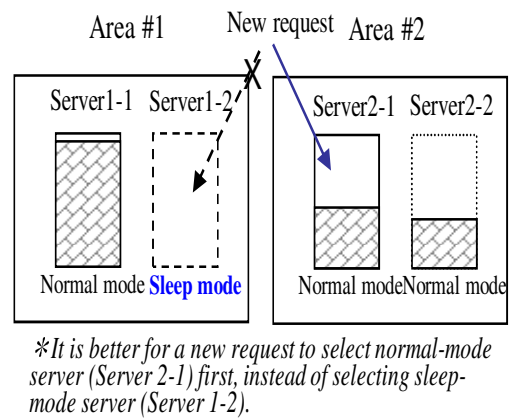

Figure 5. Example of area migration according to Policy II

Figure 6. Example of server selection according to Policy III

[1] If servers and network devices are aggregated independently without taking electric power storage into consideration, mismatches similar to Figure 4-(1) may arise and the amount of power available to servers and network devices may be reduced, as shown in Figure 7-(1). As ICT equipment cannot operate without electric power, it would be one possible approach to aggregate servers and network devices in the area which has a largest amount of available electric power capacity (**Policy IV**), as shown in Figure 7-(2). This can prolong the time during which services are provided. The possible aggregation algorithm according to policy IVis outlined as follows:

<Step 1> Area S which has the largest available electric power is selected from among $m_2$ areas. $m_2$ is the number of areas which are not in sleep mode.

<Step 2> First, we focus on processing ability.

1) Area Tc which used the least amount of processing ability is selected from among ($m_2$-1) areas.

2) If all requests currently processed in area Tc could be migrated to area S, all requests currently processed in area Tc are migrated to area S (both processing ability and bandwidth should be migrated) and area Tc gets into sleep mode  Then go back to 1). Otherwise, requests will not be migrated and stop the algorithm.

<Step 3> Then, we pay attention to bandwidth, assuming that the aggregation in step 2 is not executed.

1) Area Tb which used the least bandwidth of resource is selected from among ($m_2$-1) areas.

2) If all requests currently processed in area Tb could be migrated to area S, all requests currently processed in area Tb are migrated to area S (both processing ability and bandwidth should be migrated) and area Tb gets into sleep mode. Then go back to 1). Otherwise, requests will not be migrated and stop the algorithm.

<Step 4> If the final number of sleep areas in step 2 is equal or more than that in step 3, the aggregation executed in step2 is adopted. Otherwise, the aggregation executed in step 3 is adopted.

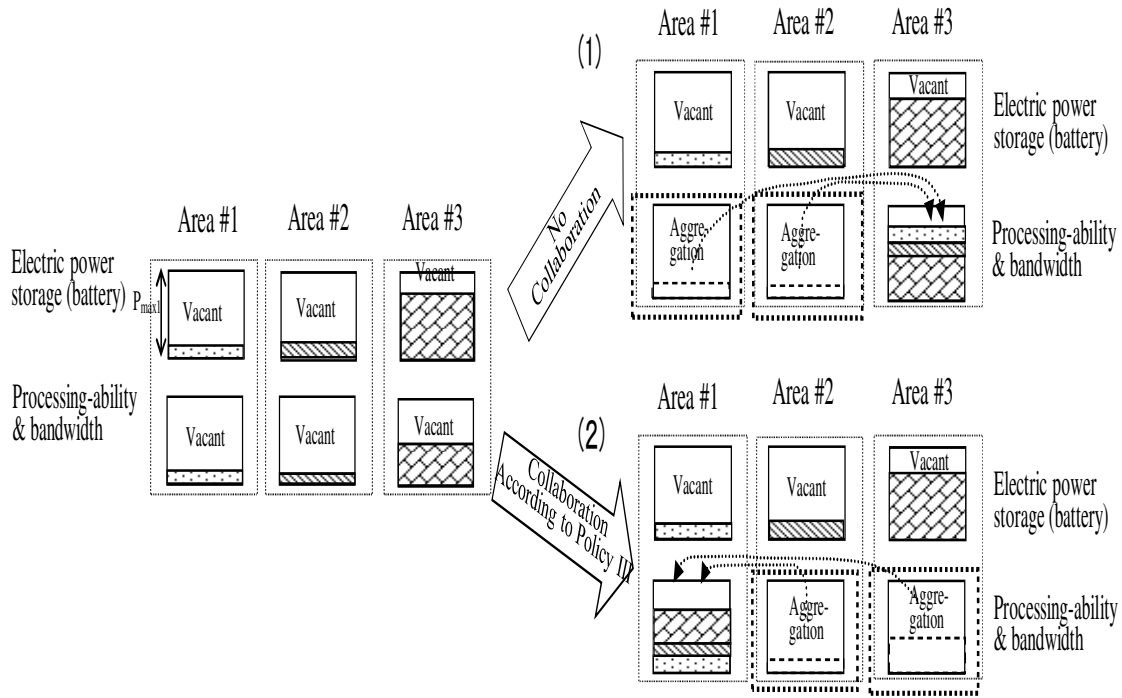

Figure 7. Collaboration image among servers, communication network and power network

Although policy IV aims to reduce the total power consumption as well as policy III, the approaches of reduction differ. That is, policy IV tries to maximize the number of requests which can be processed while policyIII tries to maximize the number of equipment of the sleep state. Policy IV is effective in the case where each area does not enough amount of electric power. Therefore, it is proposed to adopt policyIII when all areas have enough amount of available electric power and to adopt policy IV when multiple areas don't have enough amount of available electric power. The possible aggregation algorithm according to the policies III and IV is outlined as follows:

1) First, we compare the proportionate size of resource (X), comparing the size of required resource with the maximum resource size, and that of electric power (Y), comparing the size of required electric power with the remaining electric power capacity.

2) If Y is more than X, processingability and bandwidth would be aggregated in the area which has a most available electric power capacity from among areas. Otherwise, processing-ability and bandwidth would be aggregated in the area which has a least available resource from among areas.

[2] If the specific area cannot get sufficient power supply, it is impossible to provide services even if it has enough spare processing ability and bandwidth. Therefore, when the remaining power capacity drops below a certain level in one area, it is necessary to take some measures, such as allocating new requests to other areas which have a spare power capacity, or restricting any new requests (**PolicyV**) .

# 3. Possible signaling sequences to exchange information on power consumption between network and servers

Torealize Policy I in Section 2.1, it is necessary to collect information on the amount of power that will be consumed by the network and servers respectively. The following four alternative methods can be conceivedfor the collection of the information:

<Alternative 1>Power consumption per unit time by servers and that by the network are uniquely specified based only on the type of requesting application (VoIP, webaccess, file transfer etc.)

<Alternative 2>Power consumption per unit time by servers and that by the network are uniquely specifiedbased only on the bandwidth used.

<Alternative 3>Power consumption per unit time by the network is uniquely specified based on the bandwidth used, and the number of nodes traversed between the originatingand destination nodes. Power consumption per unit timeby servers is specified by the user at the time ofsubscription.

<Alternative 4>Power consumption per unit time of each service request is estimated by exchanging the information among server, client terminal and the network when a connection isestablished.

Considering the accuracy of collected information, alternative 4 would be the best solution.

Figure 8 proposes possible signaling sequences for alternative 4. The power consumption of transmission equipment between nodes is included in the power consumption by the node here. In case 1 of Figure 8, the originating client terminal A indicates the reducible volume of power consumption ($D_A$) which could be reduced with an action by terminal A. In addition, terminal A indicates the value of 'additional bandwidth holding time ($T_x$)', which will be added as a result of the action by terminal A. All nodes on the route calculates the additional power consumption by the node ($U_1$, $U_t$, $U_2$) based on Tx and transfers the calculation result toward the destination server B. The destination node (node 2) indicates the total power consumption by all nodes ($U_{12}$) on the route to server B, in addition to $D_A$ and $T_x$. Server B calculates the reducible volume of power consumption ($D_B$) based on $T_x$, which could be reducedwith an action by server B. Finally, server B judges whether the proposed action by terminal A shouldbe taken or not as follows:

If $(D_A+D_B) > U_{12}$, then take the proposed action to reduce power consumption at terminal and server, and '1' is set toflag in Connect message. Otherwise, '0' is set to flag. Server B returns Connect message toward terminal A. Terminal A and all nodes on the route can understand that the proposed action should be taken when flag is '1' . When flag is '0', any action should not be taken.

In case 2 of Figure 8, terminal A indicates the additional volume of power consumption ($U_A$) which could be increased with an action by terminal A. Ty is the value of 'bandwidth holding time' which could be decreased as a result of the action by terminal A. The very similar procedure discussed for case 1 is also appliedto case 2.

In case 3 of Figure 8, the originating node (node 1)indicates the power consumption volume ($D_1$) toward server B, which could be decreased with an action by node 1. In

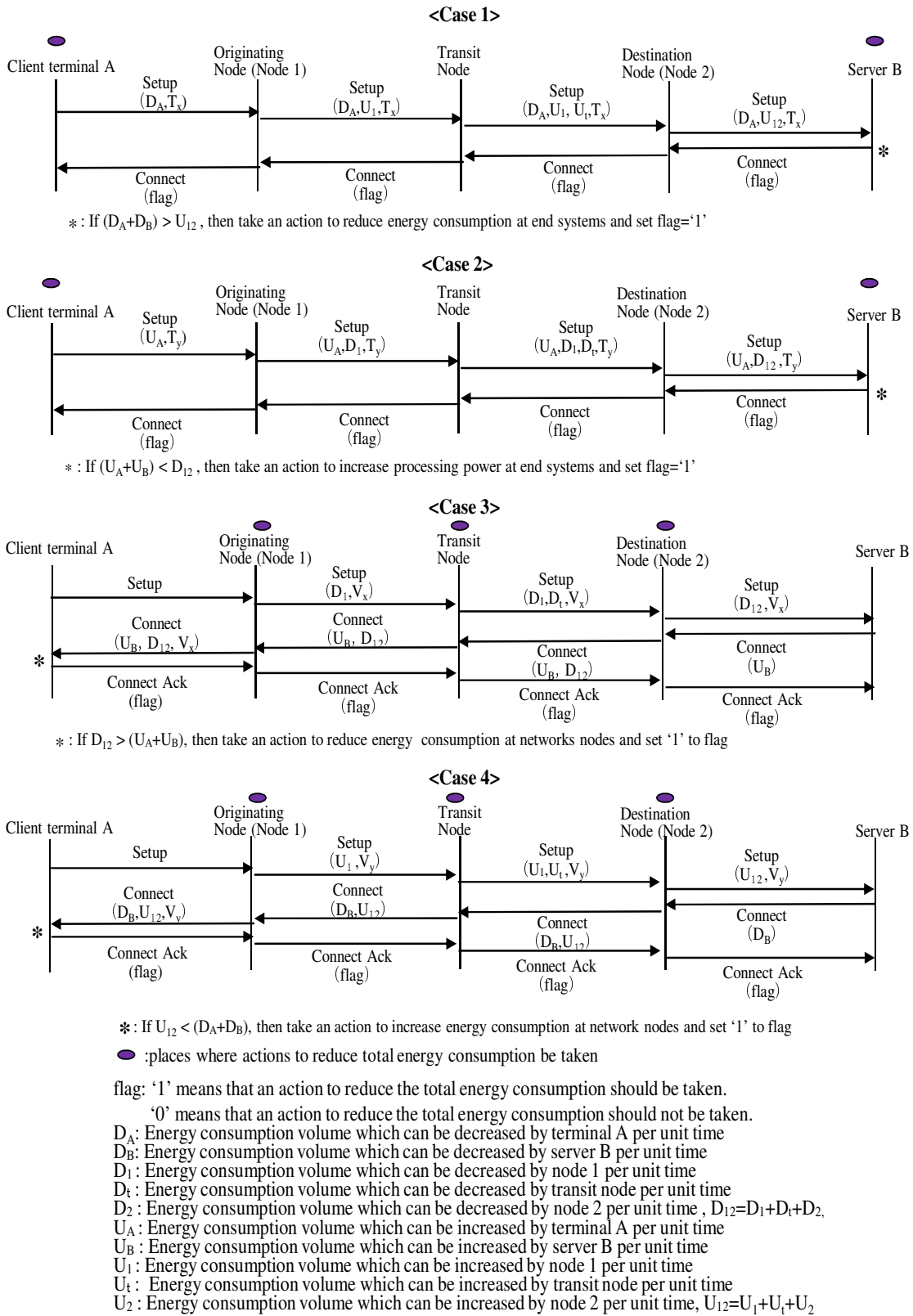

Figure 8. Possible signaling sequences for alternative 4 to collect information on energy consumption

addition, node 1 indicates the value of 'reducible bandwidth ($V_x$)' which could be reduced as a result of the action by node 1. The transit node and the destination node (node 2) calculate the reducible power consumption by the node ($D_t$, $D_2$) based on $V_x$ and transfer the calculation result toward the destination server B. Node 2 indicates the total power consumption by all nodes ($D_{12}$) on the route to server B, in addition to $V_x$. End system B calculates the additional volume of power consumption by server B ($U_B$) based on $V_x$, and judges whether the proposed actionby node 1 should be taken or not as follows:

If $D_{12} > (U_A+U_B)$, then take the proposed action to reduce power consumption at nodes and '1' is set to flag inConnect Ack message. Otherwise, '0' is set to flag.

Terminal A returns Connect Ack message with flag toward server B. Server B and all nodes on the route can understand that the proposed action should be taken when flag is is '1'. When flag is '0', any action shouldnot be taken.

In case 4 of Figure 8, the originating node (node 1) indicates the power consumption volume toward server B, which could be added with an action by node 1 ($U_1$).Vy is the value of 'additional bandwidth' which could be increased as a result of the action by node 1. The very similar procedure discussed for case 3 is alsoapplied to case 4.

Note that the information exchange proposed in the above would require complicated processing. It would be possible to eliminate the need for such information exchange by estimating the power consumed at server and terminal.Figures 9 and 10 show examples of the electric power measurementof a server that performs file transfer or handles web access respectively. If these data are stored in the network in advance, it may be possible to estimate the power of the server by measuring the volume of data transferred, or the number of transactions processed.

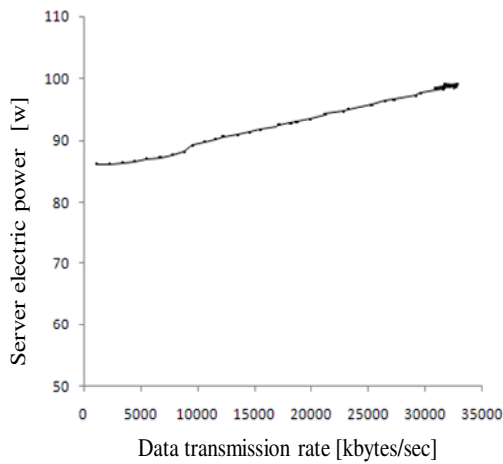

Figure 9. Data transmission rate vs. server electric power (File transfer)

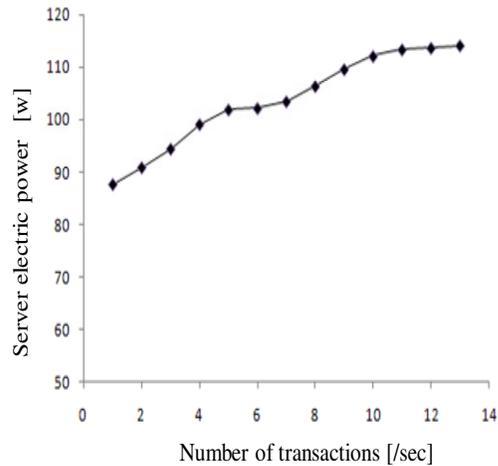

Figure 10. Number of transactions vs. server electric power (Web access)

# 4. Measurement of power consumption by all network devices and its allocation to individual user

To reduce the power consumed by the network, it is required to measure the power consumed by all network devices accurately and simply.

For transportation services, it is necessary to distribute the fuel (or power) consumed during transportation to individual cargo owner when a car carries a mix of cargoes of multiple owners. As an individual owner cannot make this allocation, the carrier could make it based on data submitted by each owner about the weight of and the distance covered by its cargo. Each cargo owner receives data on its fuel (or power) consumption from the carrier. The carrier will collect the data on power consumption by calculating consumed fuel of each truck or attaching a wireless tag to each package or managing carbon footprint, which is the history of where and how much carbon dioxidehas been emitted, for each package.

As the unit of cargo carried in the network may be a packet, a flow of packets or a established connection, it is not feasible to adopt the same way practiced in transportation system. To solve this problem, this section considers a method in which multiple probes are installed in the network to measure traffic distribution at different hours of day, and based on that data, the power consumed by the network is calculated and assigned to individual users. For the purpose, a new MIB (ex. Reference [19]) should be defined to collect data on power consumption by each network device, and a system can be constructed to collect data periodically. The information on packet routing paths in different hours of day is also necessary. The routing path of user packets in the network can be identified by sending route identifying packets (ex. tracert) between all edge nodes periodically. The ratio of power consumptionassigned to a target traffic volume is given by

$$\frac{[\text{target traffic volume}]*[\text{total energy consumption by given node}]}{[\text{total volume of traffic that goes through a given node}]} \quad (1)$$

This is calculated at different hours of day.

Figure 11 illustrates an image how to allocate power consumption atnetwork node to individual traffic flows. The entire system required for the above proposed method is illustrated in Figure 12. The monitoring system will create the traffic distribution table and the management table of packet routing path in different hours of day. Here, we suppose the network system which has m servers or terminals, m edge nodes and k transit nodes. Each server or terminal is supposed here to be accommodated to the dedicated edge node.When table 1 and table 2 are supposed to be traffic distribution table and packet routing path table respectively, the power consumption $P_{12}$ assigned to traffic volume $v_{12}$ is calculated according to Equation (1) as follows:

$$P_{12}=P_{e1}*r+P_{e2}*r+P_{t2}*v_{12}/\{v_{12}+v_{1m}+v_{23}+v_{m(m-1)}\} \quad (2)$$

$$r = v_{12}/\{(v_{11}+v_{12}+\ldots+v_{1m})+(v_{21}+v_{31}+\ldots+v_{m1})\}$$

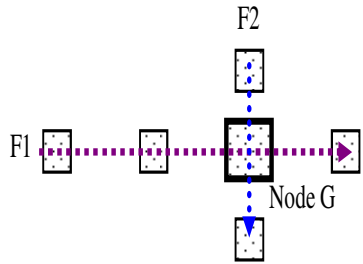

*Energy consumption of node G is divided to each flow (F1 and F2) based on traffic volume*

Figure 11. Image of allocating energy consumption at node G to individual user

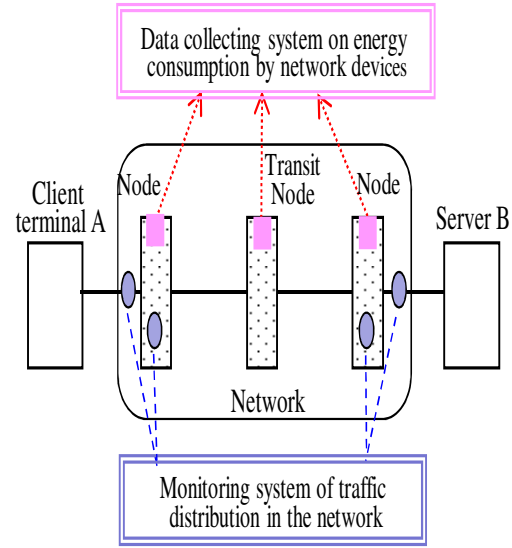

Figure 12. System for measuring and monitoring of energy consumption by network devices

Note that $P_{e1}$, $P_{e2}$, $P_{t2}$ are the total power consumption by edge node 1, edge node 2,

Table 1. Example of traffic distribution table

| Source edge node \ Destination edge node | Node 1 | Node 2 | ~ | Node m |
|---|---|---|---|---|
| Node 1 | $v_{11}$ | $v_{12}$ |  | $v_{1m}$ |
| Node 2 | $v_{21}$ | $v_{22}$ |  | $v_{2m}$ |
| ∫ | ∫ | ∫ | ∫ | ∫ |
| Node m | $v_{m1}$ | $v_{m2}$ | ~ | $v_{mm}$ |

$v_{ij}$: Traffic volume transmitted from edge node i toward edge node j

Table 2. Example of packet routing path table

| Edge node pair \ Edge or transit node | Edge node | | | | Transit node | | | |
|---|---|---|---|---|---|---|---|---|
|  | Node 1 | Node 2 | ~ | Node m | Node $t_1$ | Node $t_2$ | ~ | Node $t_k$ |
| Pair$_{1->2}$ | '1' | '1' | ~ | '0' | '0' | '1' | ~ | '0' |
| Pair$_{1->3}$ | '1' | '0' | ~ | '0' | '1' | '0' | ~ | '0' |
| ∫ | ∫ | ∫ |  | ∫ | ∫ | ∫ |  | ∫ |
| Pair$_{1->m}$ | '1' | '0' | ~ | '1' | '1' | '1' | ~ | '0' |
| Pair$_{2->1}$ | '1' | '1' | ~ | '0' | '1' | '0' | ~ | '0' |
| Pair$_{2->3}$ | '0' | '1' | ~ | '0' | '0' | '1' | ~ | '0' |
| ∫ | ∫ | ∫ |  | ∫ | ∫ | ∫ |  | ∫ |
| Pair$_{2->m}$ | '0' | '1' | ~ | '1' | '1' | '0' | ~ | '1' |
| ∫ | ∫ | ∫ |  | ∫ | ∫ | ∫ |  | ∫ |
| Pair$_{m->(m-1)}$ | '0' | '0' | ~ | '1' | '0' | '1' | ~ | '1' |

Pair $_{i->j}$: Pair of source edge node i and destination edge node j
'0': Not on the routing path, '1': On the routing path transit node $t_2$, respectively. As a result, the total power consumption $POW_g$ assigned to server g (or terminal g) is given by

$$POW_g = \sum_{i=1}^{m} (v_{gi} + v_{ig}) \qquad (3)$$

International Journal of Computer Networks & Communications (IJCNC) Vol.4, No.2, March 2012## 5. Related work

Most of conventional measures to save the power consumed by servers and the network have been discussed and implemented independently.A typical solution is to introduce power saving technologies to ICT devices and data centers [8],[9]. Proposed measures to save the power consumed by servers include not only introducing power saving technologies to them and using a virtual device concept but also slowing processor clocks or reducing the number of operating servers (turning the power of some servers off or putting them in sleep mode) while the total load on the servers is small. Power saving technologies have been also actively developed in the field of wireless LANs (IEEE802 wireless) and sensor networks [16]-[18], which areessentially driven by batteries. These technologieshave been incorporated inwireless access procedures. As for measures to save the power consumed by the network, References [10]-[14] have proposed. Representative measures proposed include putting unused devices, lines or nodes in sleep mode, controlling processor clock frequencies, controlling the line speed (rate adaptation), the number of operating lines(link aggregation) or the number of active nodes (route aggregation).

Moreover, the method to assign the power consumption by the network to an individual user had not been fully discussed.

## 6. Conclusions

In order to reduce the total power consumption by cloud computing environments, this paper first has identified the need of the collaboration among servers, communication network and power network. Five fundamental policies for the collaboration and the algorithm to realize each policy has been outlined.   Next, this paper has proposed possible signaling sequences to exchange information on power consumption between network and servers, in order to realize the proposed policy. Then, in order to reduce the power consumption by the network, this paper has proposed a method of estimating the volume of power consumption by all network devices simply and assigning it to an individual user.

In the future, it is necessary to study the detailed algorithm for realizing each collaboration policy and to clarifythe effectiveness of each policy on reducing total power consumption. It is also necessary to study the optimal location of control functions for the signaling sequences to exchange information on power consumptionbetween network and servers. As for the estimation of network power consumption, the impacts of imbalance between the volumes of upstream and downstream traffic, multipoint connections and cases where multiple networks areinvolved, are required to be studied.

### Acknowledgment

We would like to thank Mr. Kenichi Hatakeyama for hisuseful suggestionson the signaling sequences to exchange the informationon power consumption between network and servers.## References

[1] ITU Symposium on "ICTs and Climate Change" Summary Report, London, June 17&18,2008
     http://www.itu.int/dms_pub/itu-t/oth/06/0F/T060F0060090001PDFE.pdf82

[2] "Green IT Initiative in Japan", METI, Japan Oct. 2008
http://www.meti.go.jp/english/policy/GreenITInitiativeInJapan.pdf

[3] J.W.Rittinghouse and J.F.Ransone: "Cloud computing: Imprementation, management, and security", CRC Press LLC, Aug. 2009.

[4] P.Mell and T.Grance, "Effectively and securely using the cloud computing paradigm", NIST, Information Technology Lab., July 2009.

[5] P.Mell and T.Grance："The NIST definition of cloud computing" Version 15, 2009.

[6] S.Kuribayashi, "Optimal Joint Multiple Resource Allocation Method for Cloud Computing Environments", International Journal of Research and Reviews in Computer Science (IJRRCS), Vol.2, No.1, pp.1-8, Feb. 2011

[7] S.Kuribayashi, "Proposed congestion control method for cloud computing environments", International journal of Computer Networks & Communications (**IJCNC**), Vol.3, No.5, pp.161-176, Sep. 2011.

[8] M.Blackburn, "Five ways to reduce data center server power consumption", "Five ways to save server power", the green grid.   http://www.thegreengrid.org/

[9] VMware Distributed power Management (DPM)   http://www.vmware.com/products/vi/vc/drs.html

[10] M.Gupta and S.Singh, "Greening of the Internet", Proc.of ACM SIGCOMM'03, pp.19-26, Aug. 2003.

[11] C.Gunaratne, K.Christensen, S.Suen, and B. Nordman, "Reducing the Power Consumption of Ethernet with an Adaptive Link Rate (ALR)," IEEE Transactions on Computers, Vol. 57, No. 4, pp. 448-461, April 2008.

[12] F.Blanquicet, "An power efficient Internet: some ongoing work", msrSeminar08 (June 2008).

[13] S.Nedevschi, L.Popa and G.Iannaccone, "Reducing network power consumption via sleeping and rate-adaptation", Proc. 5th USENIX Symposium on Networked Systems Design and Implementation", April 2008.

[14] U.Lee, I.Rimac and V.Hilt, "Greening the internet with content-centric networking", Proceedings of the 1st International Conference on Energy-Efficient Computing and Networking (2010).

[15] E.Jung and N.H.Vaidya, "An power efficient MAC protocol for wireless LANs", Proc. IEEE INFOCOM, June 2002.

[16] X.Wu, A.Jaekel and A.Bari, "Optimal channel allocation with dynamic power control in cellular networks", International Journal of Computer Networks & Communications (**IJCNC**) Vol.3, No.2, March 2011

[17] B.Hohlt, L.Dohertly and E.Brewer, "Flexible power scheduling for sensor networks", IEEE and ACM Third International Symposium on Iformation Processing in Sensor Networks, April 2004.

[18] E.Jung and N.H.Vaidya, "An Energy efficient MAC protocol for wireless LANs", In Proc. IEEE INFOCOM, June 2002.

[19] F.Blanquicet and K.Christensen, "Managing power use in a network with a new SNMP power state MIB", IEEE Conference on Local Computer Networks (LCN) 2008, April 2008.

[20] Y.Matsumoto, S.Yanabu, "A vision of an electric power architecture for the next generation", Electrical Engineering in Japan Vol. 150, Issue 1 , pp. 18 – 25, January 2005.

[21] "Power in Japan (2008)",METI, Japan    http://www.enecho.meti.go.jp/topics/power-in-japan/english2008.pdf

[22] K.Hatakeyama, Y.Osana and S.Kuribayashi, "Reducing total power consumption with collaboration between network and servers", Proceeding of the 12-th International Conference on Network-Based Information Systems (NBiS-2009), Aug. 2009.

[23] S.Kuribayashi, "Reducing total ICT power consumption with collaboration among end systems, communication network and power network", Proceeding of the 25th IEEE International Conference on Advanced Information Networking and Applications (AINA-2011), Mar. 2011

**Author**

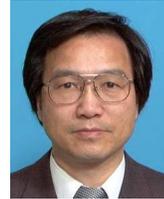

**Shin-ichiKuribayashi** received the B.E., M.E., and D.E. degrees from Tohoku University, Japan, in 1978, 1980, and 1988 respectively. He joined NTT Electrical Communications Labs in 1980. He has been engaged in the design and development of DDX and ISDN packet switching, ATM, PHS, and IMT 2000 and IP-VPN systems. He researched distributed communication systems at Stanford University from December 1988 through December 1989. He participated in international standardization on ATM signaling and IMT2000 signaling protocols at ITU-T SG11 from 1990 through 2000. Since April 2004, he has been a Professor in the Department of Computer and Information Science, Faculty of Science and Technology, Seikei University. His research interests include optimal resource management, QoS control, traffic control for cloud computing environments and green network. He is a member of IEEE, IEICE and IPSJ.